# Free-form broadband flat lenses for visible imaging

*Monjurul Meem,[1,*] Apratim Majumder[1,*] and Rajesh Menon[1,2,a)]*

[1]Dept. of Electrical & Computer Engineering, University of Utah, Salt Lake City, UT 84112, USA.

[2]Oblate Optics, Inc. San Diego CA 92130, USA.

[*]Equal contribution.

a) rmenon@eng.utah.edu

**Abstract:**

In the absence of separate apertures, the size and focal length of a lens determines both its resolution and light-collection ability defined by numerical-aperture (NA) and f/#, respectively. We utilize free-form broadband-diffractive-optical elements (BDOEs) to create flat lenses that decouple NA from f#, whereby the resolution of the lens may be chosen independently from its light-collection ability. Specifically, we designed, fabricated and characterized three BDOE lenses operating in the visible band (450nm-750nm), each with f/11.25, but with NAs of 0.00075, 0.0067 and 0.054, respectively. Simulations confirm that such decoupling is possible even at much higher NAs. Experiments confirm achromatic focusing and broadband imaging. One of the lenses exhibited a depth-of-focus almost 2 orders of magnitude larger than the diffraction limit. Such BDOE lenses could be very useful in focal-plane arrays with large pixel sizes, where light collection efficiency needs to be maintained. Furthermore, by abandoning rotational symmetry, one can achieve free-form geometries in the focal spot, such as a square that can more closely match the geometry of the sensor pixel.

**Introduction:** The f-number (f/#) of a lens is a measure of its light gathering power, which directly controls the image brightness. The f/# is defined as the ratio of the focal length (f) of the lens to the entrance pupil diameter (D). The f/# can be adjusted by tuning the entrance pupil diameter, so decreasing the f/# will increase the area of the aperture, effectively increasing the light throughput or light-collection ability [1]. But, varying the f/# alters the numerical aperture (NA) of the lens as well, which in turn affects its resolution [1]. The NA of a lens is defined as the multiplication of the refractive index (n) of the medium in which light propagates and sine of the half of the maximum angle (2θ) with which the light is focused into the focal spot. For a lens operating in air (n = 1), NA can be written as [2]

$$NA = \sin\theta = sin[\tan^{-1}(\frac{D}{2\times f})] \approx \frac{D}{2\times f} = \frac{1}{2\times (f/\#)}, \tag{1}$$

where the approximation is valid for small NA. It is important to note that f# and NA cannot be changed independently in a single lens with a separate aperture. This poses a unique challenge, for example, if we want to keep large light collection ability, while trying to create a large focal spot. This is advantageous for example, when the sensor pixel size is much larger than the diffraction limit [1,4].

The NA of lens is a measure of how tightly the light is focused. The smallest extent of the focused spot, which is also referred to as the point-spread function (PSF) is limited by diffraction as:

$$NA_{eff} = \frac{\lambda}{2\times S}, \tag{2}$$

where S is the full-width at half-maximum (FWHM) of the PSF at wavelength, λ. We note that the focal spot can, in fact, be larger than this limit (up to the aperture size of the lens itself). So, by engineering the size and geometry of the focus spot, one can effectively decouple f# from NA.

This can be achieved readily by engineering the complex transmittance of the lens, also referred to as the lens pupil function. Engineering the pupil function has enabled super-resolution microscopy [5] and lithography [6,7], extended depth of field [8,9] and achromaticity across extreme bandwidths [10-12]. Modulation of the pupil function can be accomplished via a spatial light modulator (SLM) [5,13], via deformable mirrors [14] or via diffractive optical elements (DOEs) [6-12]. DOEs have the advantages of large space-bandwidth product, large range in phase, very low-loss, efficient broadband operation and potential to handle high power [15].

Here we demonstrate three broadband DOE-based flat lenses operating in the visible spectrum ($\lambda$ = 450nm to 750nm). All 3 flat lenses have square apertures with side 4mm and focal length of 45mm (Fig. 1a), which yields a geometric f# of 11.25, but by engineering the spot sizes to be 400μm, 45μm and 5.6μm, we were able to achieve effective NAs of 0.00075, 0.0067 and 0.054, respectively (Fig. 1b). Our approach is not limited to small NAs and we showcase two simulated BDOE lenses with f/# of 2.5, but effective NAs of 0.3 and 0.03 illustrated in Figs. 1(c-f).

**Design:** Previously we have demonstrated the concept of BDOE as a lens, whose point spread function (PSF) is structured in the form of an image [16,17]. To avoid confusion from non-imaging devices (which have sometimes been misidentified as lenses), by a lens we mean that the lens-makers equation is satisfied by the object and image distances, and images can be formed. The BDOE is comprised of square pixels, whose minimum width is determined by the fabrication technology, and whose heights are determined via a nonlinear optimization procedure that has been described previously [15,16]. The goal of the optimization is to maximize the diffraction efficiency based on the target image, averaged over all the wavelengths of interest. In this work, the target image is a focused spot, whose geometry and size are selected independently of the f#

of the lens. Specifically, we designed, fabricated and characterized three BDOE lenses to focus light into a square spot of width = 400µm (L1), 45µm (L2) and 5.6µm (L3) respectively as illustrated in Fig. 1(a,b). For simplicity, we constrained the pixel size to be 5µm with 100 possible height levels between 0 and 2.6µm. Assuming an average wavelength of 600nm and designed spot sizes of 400µm, 45µm, 5.6µm, the effective NA of the BDOE lenses are 0.00075, 0.0067 and 0.054, respectively. The optimized height profiles along with the simulated point-spread functions of the BDOE lenses are shown in Fig. 1(b).

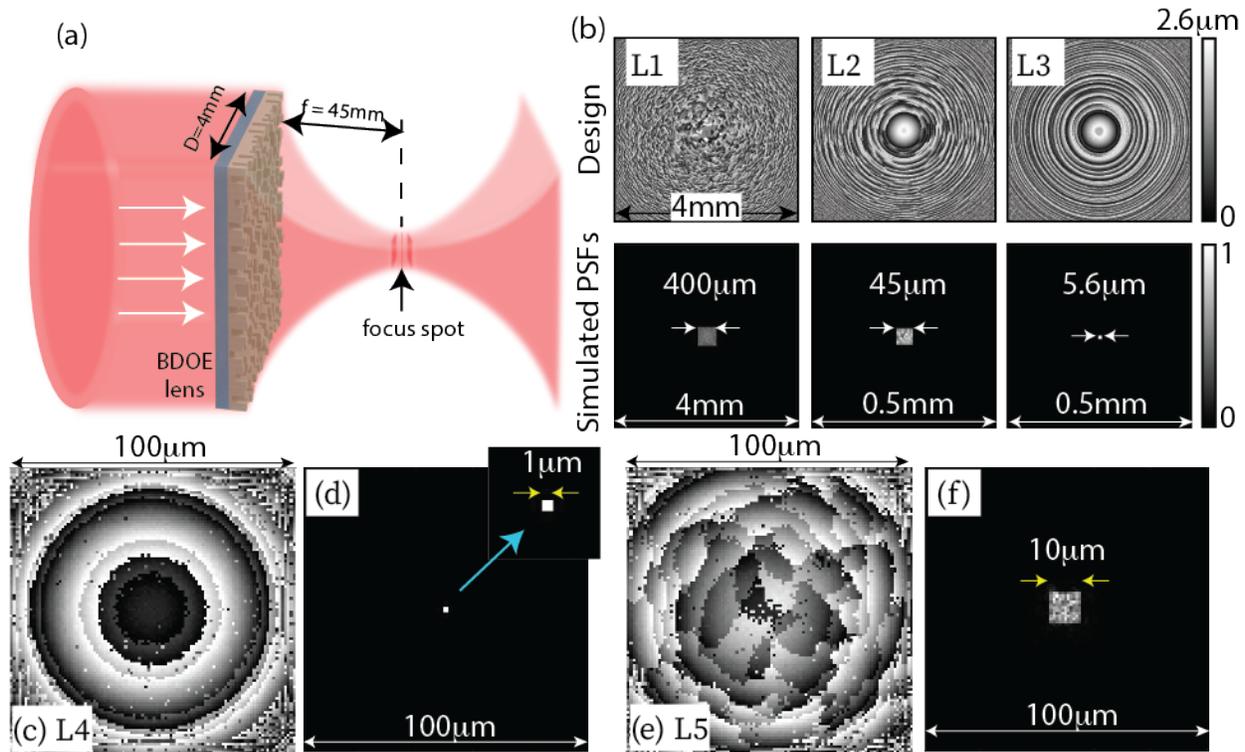

**Figure 1:** *(a) Free-form flat lens with square aperture, side=D=4mm and focal length, f=45mm. The constituent pixels are squares of side 5µm. (b) Top-row: Optimized pixel height distributions with spot-size of 400µm (L1), 45µm (L2) and 5.6µm (L3). Bottom row: Simulated point-spread functions. Note the square focal spots. Flat-lenses with f=250µm, size=100µm (f/2.5), and (c) spot-size=1µm (L4, $NA_{eff}$=0.3) and (e) spot-size=10µm (L5, $NA_{eff}$=0.03). Corresponding visible band PSFs in (d) and (f), respectively. Color axis is same for (c) and (e) as top-row of (b). Color axis is same for (d), (f) as bottom-row of (b).*

For ease of fabrication, we limited our experiments to low $NA_{eff}$. However, this is not a fundamental limitation as illustrated by two flat lens designs showcased in Figs. 1(c-f) with

f/#=2.5, but $NA_{eff}$ of 0.3 (L4) and 0.03 (L5). All design parameters were the same as in L1-L3, except pixel width was 1μm, the number of pixels was 100 X 100 (lens aperture size = 100μm), and focal length = 250μm. Clearly, engineering the spot-size and thereby, decoupling f/# and NA is possible at higher $NA_{eff}$ as well.

**Experiments and Discussion:** The BDOEs were fabricated by patterning a positive-tone, transparent photopolymer film (S1813, Microchem), which was spun-cast to a thickness of 2.6μm on top of a soda-lime-glass substrate (thickness~0.5mm, diameter=50mm). Laser gray-scale lithography was used for patterning [9-11]. Optical micrographs of the fabricated devices are shown in Fig. 2.

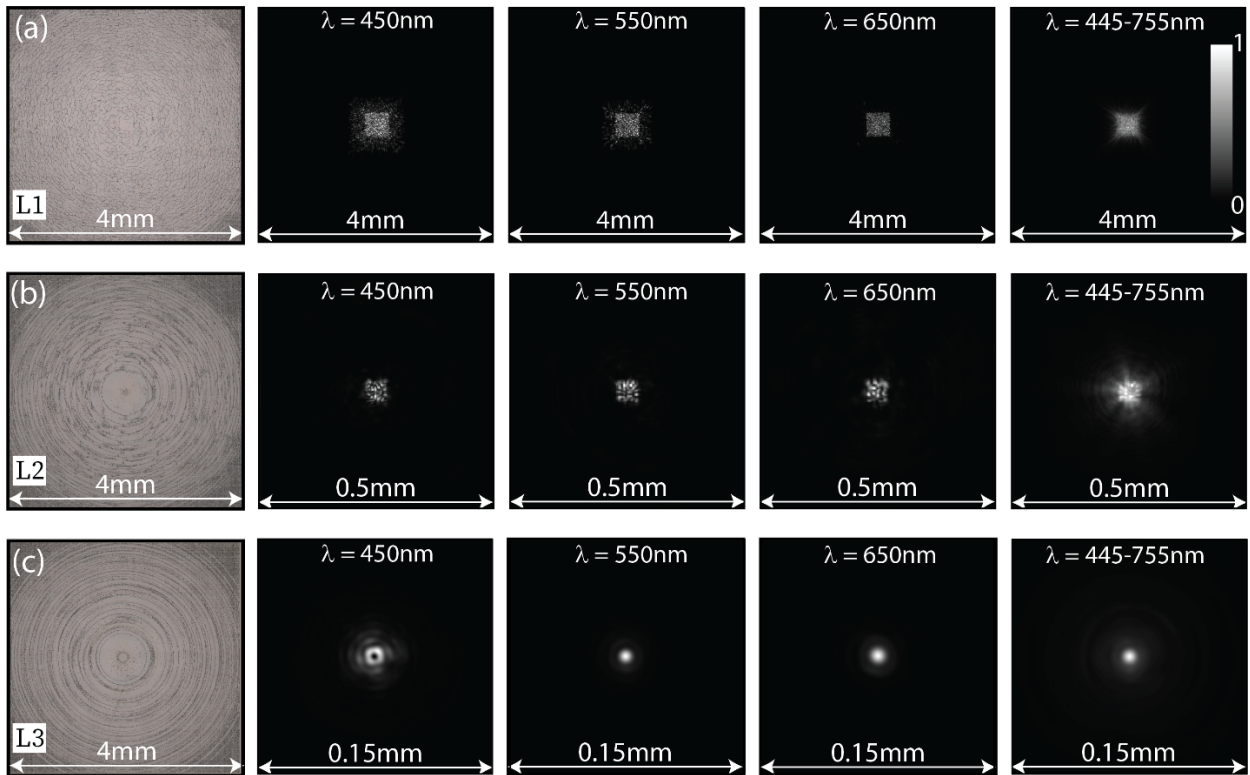

*Figure 2: Experimental verification of broadband focusing of (a) L1, (b) L2 and (c) L3 lenses. From left to right: Optical micrographs of the fabricated lenses, Measured PSFs for the narrowband (450nm, 550nm, 650nm) and broadband illuminations (455-755nm). The illumination bandwidth was 15nm for the narrowband experiments.*

The focusing performance of each flat lens was characterized by recording the point-spread function (PSF) under broadband and narrowband wavelengths. Each lens was illuminated by an expanded and collimated beam from a super-continuum source (SuperK EXTREME EXW-6, NKT Photonics) coupled to a tunable filter (SuperK VARIA, NKT photonics). The PSF was recorded directly on a monochrome CMOS image sensor (DMM 27UP031- ML, The Imaging Source). The captured PSFs under narrowband primary colors (red, green, blue) illumination and broadband white light illumination are shown in Fig. 2. Achromatic focusing with spot size close to that predicted by simulation is observed (see supplementary information), although some of the discrepancies can be attributed to the fabrication errors [16]. For characterizing the depth of focus (DOF), we placed the image sensor on a stage and captured the PSFs at different distances under broadband illumination (455 - 755nm). The results are illustrated in Fig. 3. Our experiments confirm a DOF of ~20mm for L3, which is somewhat larger than that expected from diffraction.

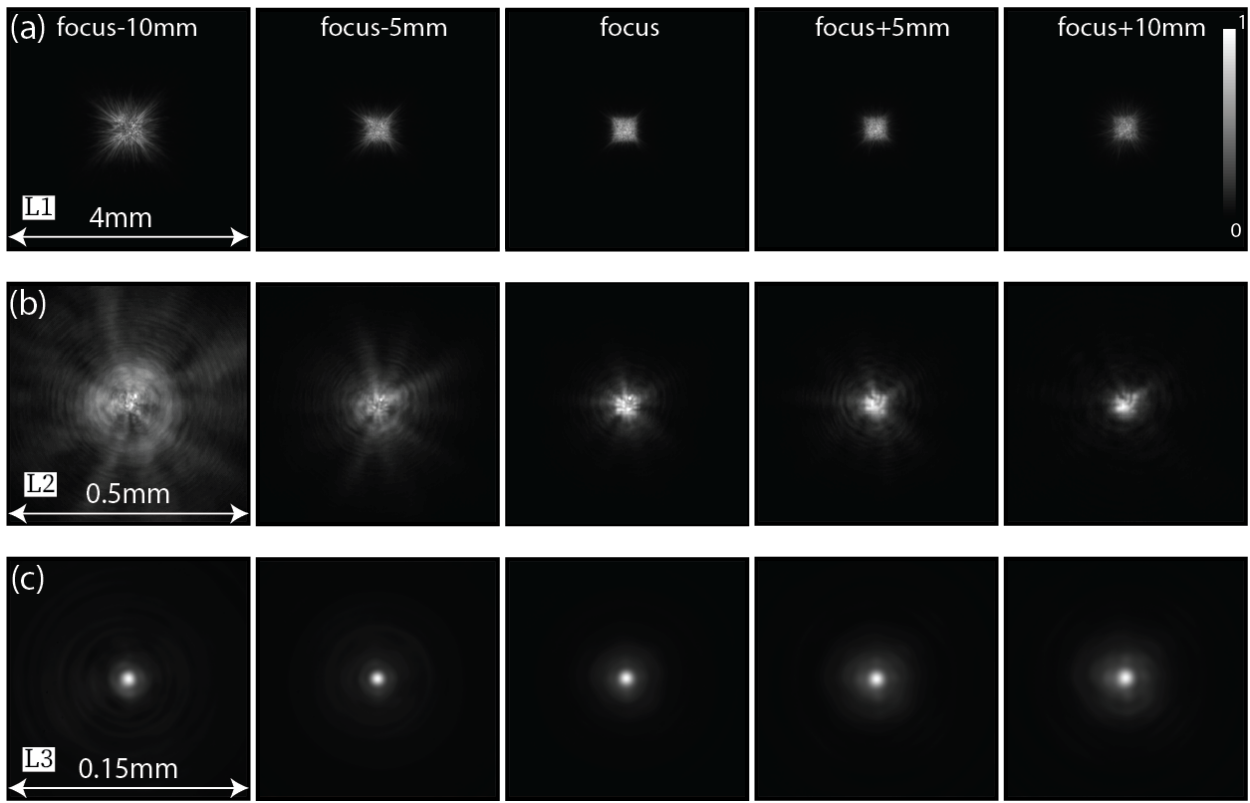

*Figure 3:* Measured PSFs as function of defocus for (a) L1, (b) L2, and (c) L3 lenses. The sensor was placed behind the lens and moved to capture the focus spot at different distances to estimate the DOF. Note focus is located at 45mm. Strehl ratio and on-axis intensity vs defocus are included in supplementary information.

Finally, we assembled a camera by placing the BDOE lens in front of a color image sensor (DFM 72BUC02-ML, The Imaging Source) and recorded still and video images as summarized in Fig. 4 (Supplementary Videos 1-3 for L1-L3, respectively). The resolution chart was back illuminated by broadband (455-755nm) white light from the super-continuum source and the color objects were front-illuminated by white LED light. The exposure time was adjusted to ensure that the frames were not saturated. In addition, dark frame was recorded and subtracted from each image. Note that, all the color images have similar magnification. The resolution chart reveals the smallest resolved spatial frequencies as 0.28 lp/mm, 11.30 lp/mm and 64 lp/mm, for L1-L3, respectively. Each lens produced images at approximately the same exposure time.

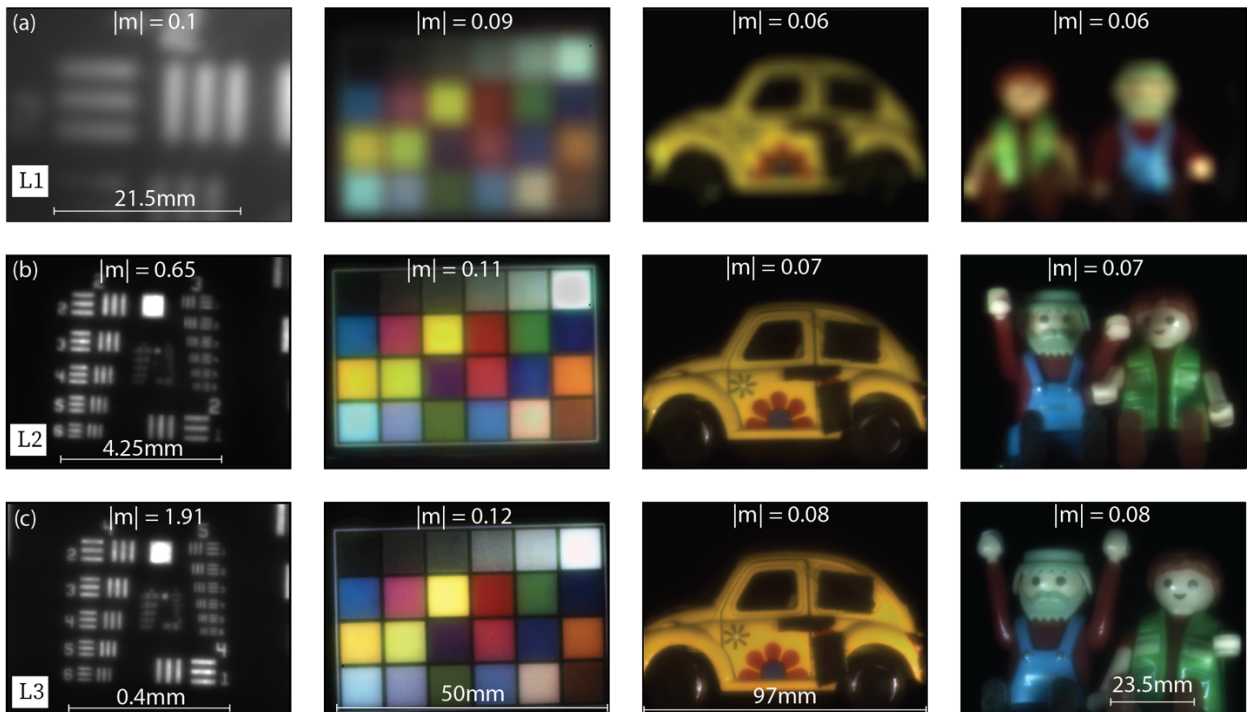

*Figure 4:* Characterization of Imaging with lenses (a) L1, (b) L2 and (c) L3 (See supplementary videos 1-3). The resolution chart was back-illuminated by the super-continuum source (455-755nm) and the color objects were front-illuminated by a white LED. The magnification (|m|) of

*each image is noted. The scale bar at the bottom represents object size. The illumination condition and exposure times were the similar for all lenses, confirming that all 3 have the same f#.*

As indicated in Fig. 3, L3 is expected to have an extended DOF. In order to confirm this, we placed L3 at a fixed distance from the sensor (70mm) and adjusted the object distance from ~70mm to 400mm. Supplementary video 4 confirms that the image remains in focus over most of this range, indicating a dramatic increase in depth of field across the visible band.

**Conclusion:** It is thought that the f# and NA of a lens are inversely related and cannot be changed independently. Here, we show that this limitation can be overcome by engineering the geometry of the focal spot via free-form BDOE flat lenses. In this paper, we experimentally demonstrated three such lenses, where we have successfully decoupled the f# and NA relationship by controlling the focus spot size. All our experimental results confirm excellent focusing and imaging performance over a broad wavelength range. Such BDOE lenses could be very useful in focal-plane arrays with large pixel sizes, where light collection efficiency needs to be maintained. Furthermore, by abandoning rotational symmetry, one can achieve free-form geometries in the focal spot, such as a square that can more closely match the geometry of the sensor pixel.

**Acknowledgements**

We thank Brian Baker, Steve Pritchett for fabrication advice

**Funding**

Office of Naval Research grant N66001-10-1-4065.

**Author Contributions**

R.M. and M.M. conceived and designed the experiments, modeled and optimized the devices. M.M. fabricated the devices. M.M and A.M. performed the experiments and numerical analysis. All authors analyzed the data and wrote the paper.

**Competing Interests**

R.M. is co-founder of Oblate Optics, which is commercializing the subject technology. The University of Utah has applied for a patent covering the subject technology.

# Supplementary Information

Free-form broadband flat lenses for visible imaging

Monjurul Meem[1], Apratim Majumder[1], and Rajesh Menon[1,2,*]

[1]Department of Electrical and Computer Engineering, University of Utah, Salt Lake City UT 84112

[2]Oblate Optics, Inc., 13060 Brixton Place, San Diego CA 92130

*rmenon@eng.utah.edu

## 1. Design and geometric parameters of the holograms

The flat lens is pixelated in X and Y directions, height of each pixel is quantized into multiple levels. Different parameters have been used for the lens designs. These are summarized in Table S1.

*Table S1. Design and geometric parameters of fabricated flat lens.*

| Flat lens | Pixel size (μm) | Number of pixels | Physical size (mm X mm) | Maximum height (μm) | Focal length (mm) | NA | F# | Operating wavelengths (nm) |
|---|---|---|---|---|---|---|---|---|
| L1 | 5 | 800 x 800 | 4 x 4 | 2.6 | 45 | 0.00075 | 11.25 | 450 - 750 |
| L2 | 5 | 500 x 500 | 4 x 4 | 2.6 | 45 | 0.0067 | 11.25 | 450 - 750 |
| L3 | 5 | 500 x 500 | 4 x 4 | 2.6 | 45 | 0.054 | 11.25 | 450 - 750 |
| L4 | 1 | 100 x 100 | 0.1 x 0.1 | 2.6 | 0.25 | 0.3 | 2.5 | 450 - 650 |
| L5 | 1 | 100 x 100 | 0.1 x 0.1 | 2.6 | 0.25 | 0.03 | 2.5 | 450 - 650 |

## 2. Fabrication

The flat lenses were fabricated using direct laser write grayscale lithography [1]. A positive tone photoresist (s1813) [2] was spin coated on a 2" soda lime glass wafer at 1000 rpm for 60 seconds to yield a thickness a 2.6 μm. The spin coated sample was baked in an oven at 110°C for 30 minutes. The designs were written on the sample using "Heidelberg μPG 101" tool [3] and developed in AZ 1:1 solution for 40 seconds. A calibration step was performed beforehand to determine the exposed depths at a particular gray scale level. The details of the fabrication process have been discussed elsewhere [4].

## 3. Focal spot characterization

The flat lens was illuminated with expanded and collimated beam from a SuperK EXTREME EXW-6 source (NKT Photonics). The wavelength and bandwidth was tuned with SuperK VARIA filter (NKT Photonics). The focus spot of the lens L1 and L2 were captured directly on a monochrome sensor (DMM 27UP031-ML, Imaging Source). For lens L3, the focal was magnified 22.2 times using a 20x objective (Olympus PLN 20X Objective) with the NA of 0.4 and tube lens (ITL200, Thorlabs). The exposure time of the sensor was carefully adjusted to avoid pixel saturation. A dark frame was also captured with all the light source turned off. A bandwidth of 15nm was used for narrowband illumination and a spectrum of 445nm to 755nm was used for broadband illumination.

For the depth of focus measurement, the sensor was placed on a stage and the PSFs was captured at 35mm, 40mm, 45mm (focal length), 50mm, 55mm away from the lens under broadband illumination.

After capturing the focal spot, the Focusing efficiency was then calculated using the following equation: Focusing efficiency = (sum of pixel values in 3*FWHM) / (sum of pixel values in the entire lens area)

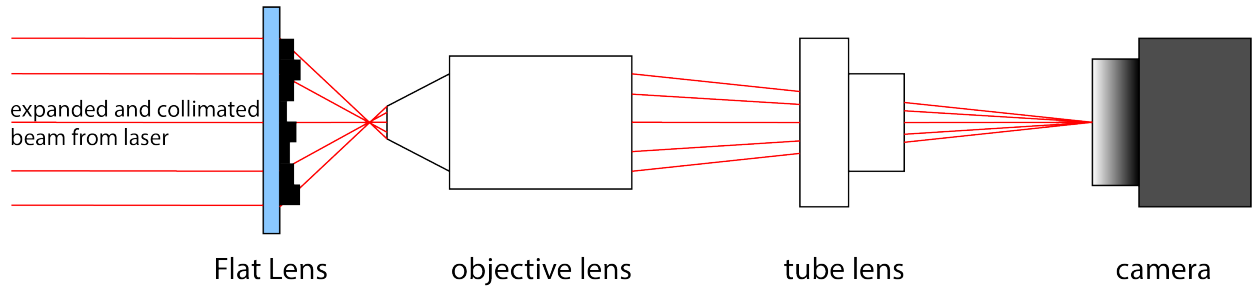

*Figure S1:* *Schematic of the set-up used to capture the focal spots of the flat lens L3. An expanded and collimated laser beam is focused by the flat lens. An objective lens paired with a tube lens is used to form a magnified image of the focal spot on a CMOS camera.*

### 4. Imaging setup

For imaging, colorful objects, the objects were placed in front of the lens and were illuminated with white LED flash light and the corresponding images were captured using a color sensor (DFM 72BUC02-ML, Imaging Source).

For resolution chart imaging, the 1951 USAF resolution test chart (R3L3S1N, Thorlabs) was used as object. The USAF target was illuminated with broadband (445nm-755nm) laser and corresponding images were captured using a monochrome sensor (DMM 27UP031-ML, Imaging Source). All the color images were taken with similar exposure setting of the sensor (1/4 sec to 1/6 sec). In each case, a dark frame was recorded and subtracted from the images.

Additionally, to prove EDOF imaging, the sensor was kept fixed at 70mm away from the lens L3 and the object was moved from 50mm to 400mm, and the corresponding video was captured. The distance was same for all the videos, for the "U sign on the mountain", it was a still image displayed on cell phone screen.

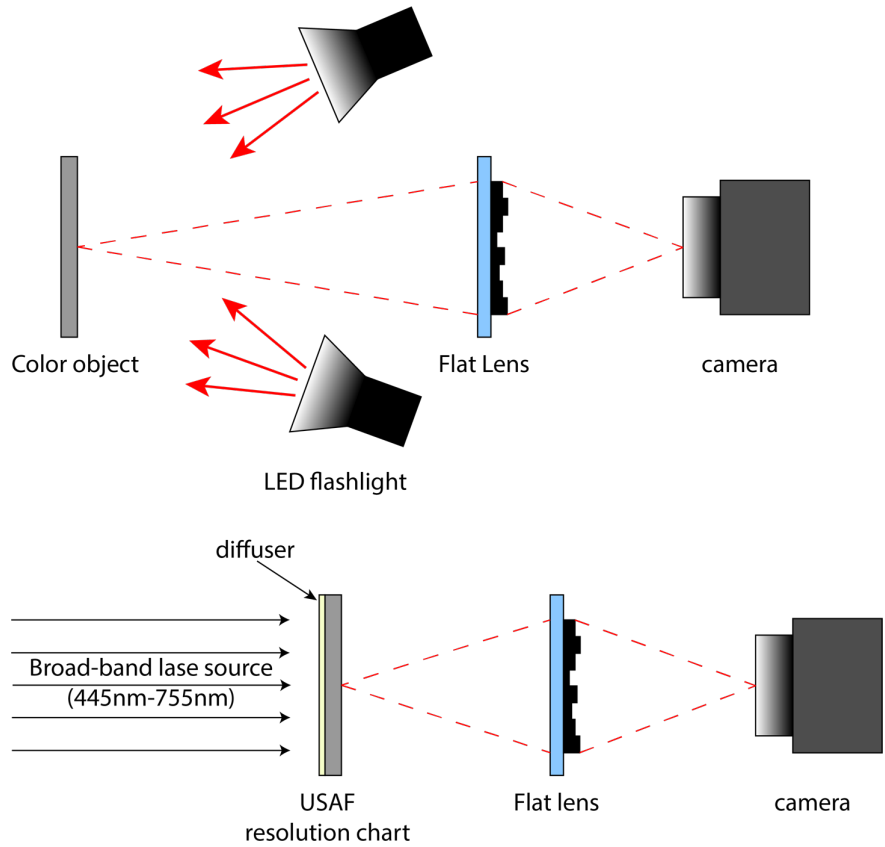

***Figure S2:*** *Schematic of the imaging setup. The object was illuminated with LED flash light or broadband laser, and the corresponding image formed by the flat lens was captured on a camera.*

## 4. Focusing performance of flat lens

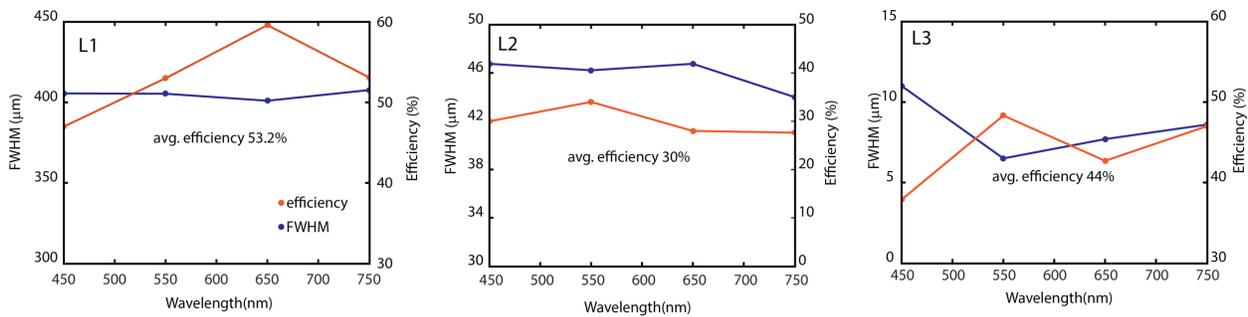

***Figure S3:*** *The measured full width at half maximum (FWHM) along with efficiency for lens L1 and L2 and L3 are given.*

## 5. Supplementary Videos:

- Supplementary Video 1 shows imaging using L1 of different objects under white LED light.
- Supplementary Video 2 shows imaging using L2 of different objects under white LED light.

- Supplementary Video 3 shows imaging using L3 of different objects under white LED light.
- Supplementary Video 4 shows imaging using L1 of different objects with fixed image distance (70mm), but object distance varying from ~70mm to 400mm.